%% file: HIN-25-015_temp.tex
\documentclass[11pt,twoside,a4paper,cmspaper,final,collab]{cms-tdr}
\def\svnVersion{3ade612-D}\def\svnDate{2026/07/14}\def\cmsCernNoTag{CERN-EP-2026-188}\def\cmsCernDate{\today}\def\cmsMessage{Submitted to Physical Review Letters}
\begin{document}\cmsNoteHeader{HIN-25-015}

\newcommand{\pp}{\ensuremath{\Pp\Pp}\xspace}
\newcommand{\pPb}{\ensuremath{\Pp\text{Pb}}\xspace}
\newcommand{\OO}{\ensuremath{\text{OO}}\xspace}
\newcommand{\NeNe}{\ensuremath{\text{NeNe}}\xspace}
\newcommand{\PbPb}{\ensuremath{\text{PbPb}}\xspace}
\newcommand{\AAa}{\ensuremath{\text{AA}}\xspace}
\newcommand{\Npart}{\ensuremath{N_{\text{part}}}\xspace}
\newcommand{\Ntrkcor}{\ensuremath{N_{\text{trk}}^{\text{corr}}}\xspace}
\newcommand{\ANpart}{\ensuremath{\langle\Npart\rangle}\xspace}
\newcommand{\DBTwoOne}{\ensuremath{D_{21}}\xspace}
\newcommand{\DBThreeOne}{\ensuremath{D_{31}}\xspace}
\newcommand{\DBThreeTwo}{\ensuremath{D_{32}}\xspace}
\newcommand{\SNTwoOne}{\ensuremath{S_{21}}\xspace}
\newcommand{\SNThreeOne}{\ensuremath{S_{31}}\xspace}
\providecommand{\SHINCHON}{\textsc{shinchon}\xspace}
\ifthenelse{\boolean{cms@external}}{
  \providecommand{\cmsLeft}{upper\xspace}
  \providecommand{\cmsRight}{lower\xspace}
  \providecommand{\enn}{\ensuremath{n}\xspace}
  \providecommand{\emm}{\ensuremath{m}\xspace}
  \providecommand{\enm}{{\ensuremath{nm}}\xspace}
  }{
    \providecommand{\cmsLeft}{left\xspace}
    \providecommand{\cmsRight}{right\xspace}
     \providecommand{\enn}{\ensuremath{\text{n}}\xspace}
    \providecommand{\emm}{\ensuremath{\text{m}}\xspace}
    \providecommand{\enm}{\ensuremath{\text{nm}}\xspace}
  }

\cmsNoteHeader{HIN-25-015}
\title{Evidence for sequential \texorpdfstring{\PGUPnS{}}{Upsilon(nS)} suppression in light ion collisions}

\author[cern]{The CMS Collaboration}
\date{\today}
\abstract{
Bound states of heavy quark-antiquark pairs, known as quarkonia, have long been regarded as particularly sensitive probes of the quark-gluon plasma (QGP). 
Comparing quarkonium yields in collisions of heavy nuclei, such as gold or lead, with a proton-proton (\pp) reference reveals a characteristic pattern of sequential suppression, in which weakly-bound excited states are more strongly suppressed than the ground states. We report the first measurements of the three lowest mass $\mathrm{S}$-wave vector bottomonium resonances, the ground state \PgUa and the excited states \PgUb and \PgUc, in oxygen-oxygen collisions at a center-of-mass energy per nucleon pair of $\sqrtsNN=5.36\TeV$. Measurements of the yields of the \PgUa and \PgUb resonances in neon-neon collisions, at the same $\sqrtsNN$, are also presented. The \PgUb{}/\PgUa ratio is found to be significantly below the measured \pp reference value, and the \PgUc{}/\PgUa ratio shows an even larger reduction. The significance of the relative \PgUc to \PgUb suppression exceeds three standard deviations. These results provide evidence for the sequential suppression of \PGUPnS{} states in light ion collisions, similar to observations made in lead-lead and gold-gold collisions, generally attributed to the presence of a QGP medium.}

\hypersetup{
pdfauthor={CMS Collaboration},
pdftitle={Evidence for sequential Y(nS) meson suppression in light ion collisions},
pdfsubject={CMS},
pdfkeywords={CMS, light ions, Upsilon, Upsilon sequential suppression, QGP}}

\maketitle

Quarkonia, bound states of a heavy quark with its antiquark~\cite{Ellis:2025nkm}, have long served as sensitive probes of the temperature and space-time evolution of the quark-gluon plasma (QGP)~\cite{Harris:1996zx}, a deconfined state of quarks and gluons. Their different binding energies and spatial sizes make individual states respond differently to the QGP~\cite{Matsui:1986dk}.  Modern descriptions of this effect include multiple mechanisms that can break up or recreate quarkonia as the QGP evolves~\cite{Andronic:2025jbp,Apolinario:2022vzg,Thews:2000rj,Braun-Munzinger:2000csl,Du:2017qkv,Yao:2020xzw,Brambilla:2023hkw,Wu:2025lcj}.
 Theoretical models incorporating these processes provide a broadly consistent interpretation of quarkonium measurements in nucleus-nucleus (\AAa) collisions with heavier elements such as gold and lead~\cite{Bahavar:2014jza,Andronic:2022ucg,Andronic:2024oxz,Wu:2025lcj}. The bottomonium family consists of bottom quark-antiquark states, the three lightest $\mathrm{S}$-wave vector ones being \PgUa, \PgUb, and \PgUc, which  have binding energies of 1.10, 0.54, and 0.20\GeV, respectively~\cite{ParticleDataGroup:2024cfk}. This variation makes the \PGUPnS family a useful differential probe of QGP properties such as temperature.  The observed hierarchy in the modification of the \PGUPnS yields relative to  proton-proton (\pp) data, together with the loss of feed-down from excited states to the ground state, is referred to as sequential \PGUPnS meson suppression.

An indication of sequential suppression of \PgU mesons in heavy ion collisions was first reported by the CMS Collaboration~\cite{CMS:2024krd}, 
which found that the $[\PgUb+\PgUc]\,/\,\PgUa$ excited-to-ground-state yield ratio
is smaller in \PbPb collisions at $\sqrtsNN = 2.76\TeV$ than in a \pp reference~\cite{CMS:2011all}. The CMS Collaboration later measured the individual \PgU states in \PbPb\ collisions at $\sqrtsNN = 2.76$ and $5.02\TeV$, finding in particular that the \PgUb/\PgUa ratio is smaller in \PbPb than in \pp by more than five standard deviations~\cite{CMS:2012gvv,CMS:2018zza}. The three-state suppression pattern was established when CMS reported the first observation of the \PgUc yield in \PbPb collisions at $\sqrtsNN = 5.02\TeV$~\cite{CMS:2023lfu}. A stronger suppression of \PgUb relative to \PgUa in \PbPb\ collisions at $\sqrtsNN = 5.02\TeV$ has also been reported by the ALICE and ATLAS Collaborations~\cite{ALICE:2020wwx,ATLAS:2022exb}. At RHIC, the STAR Collaboration has also measured sequential \PgU suppression in gold-gold collisions at $\sqrtsNN = 200\GeV$~\cite{STAR:2022rpk}. Clear signs of \PgU suppression have also been reported in the much smaller proton-lead (pPb) system at the LHC~\cite{ALICE:2019qie,ATLAS:2017prf,CMS:2022wfi,LHCb:2018psc}. These observations raised the question of whether quarkonium suppression patterns established in collisions of heavier ions retain the same interpretation across collision systems. On the one hand, the short lifetime and small space-time extent of the \pp and \pPb  collision systems provide sensitivity to early-time quarkonium dynamics~\cite{Emerick:2011xu,Bagchi:2023vfv}. On the other, the \pPb measurements have motivated intense theoretical debate over the contribution of final-state effects, including quarkonia dissociation by co-moving particles and a short-lived QGP~\cite{Sahoo:2019ifs,Gardim:2022yds,Ke:2022gkq,PHENIX:2023dxl,FerreiroComp,CMS-PAS-HIN-25-005}. 
 
However, experimentally disentangling cold nuclear matter (CNM) effects from final-state interactions remains challenging in \pPb collisions. Therefore, a better understanding of the mechanisms responsible for the observed sequential \PgU suppression requires studying systems of varying sizes and whose initial conditions differ from those in \pPb. 
Collisions of lighter ions, such as oxygen and neon, can provide this controlled intermediate size regime. 

Oxygen-oxygen (\OO) and neon-neon (\NeNe) collisions were performed at the LHC in 2025~\cite{Waagaard:2025dsi}, providing an opportunity to test QGP-sensitive observables, including bottomonium suppression, in symmetric nuclear collision systems an order of magnitude smaller than \PbPb~\cite{Brewer:2021kiv,ALICE:2025luc,ATLAS:2025nnt,CMS:2025tga,CMS:2025bta,CMS:2026qef,LHCb:2026gwz,CMS:2026sai,ALICE:2026zck,ATLAS:2026zgq,ATLAS:2026fos}. The collisions were performed at $\sqrtsNN=5.36\TeV$, enabling  direct comparison with a \pp reference data sample collected at the same energy in 2024.  In this Letter, we report the first measurements of the \PgUa, \PgUb, and \PgUc states in \OO collisions, and of the \PgUa and \PgUb states in \NeNe collisions. The data were collected with the CMS detector, corresponding to  total integrated luminosities of $8.4\pm0.3\nbinv$, $0.81\pm0.03\nbinv$, and $456\pm14\pbinv$ for the \OO, \NeNe, and reference \pp samples, respectively~\cite{CMS:2025rzq}. The suppression is quantified through  double ratios of excited-to-reference-state \PgU yields in \AAa collisions relative to \pp collisions~\cite{CMS:2011all}:
\begin{equation*}
D_\enm
= \Bigg( \frac{N^{\PGUPnS}}{N^{\PGUP{mS}}} \Bigg)_{\AAa}
   \Bigg/
  \Bigg( \frac{N^{\PGUPnS}}{N^{\PGUP{mS}}} \Bigg)_{\pp}.
\end{equation*}
Here, \enn and \emm denote the two \PgU states entering the ratio, with $(\enn,\emm)=(2,1),(3,1)$, and $(3,2)$. The yield ratio that forms either the numerator or the denominator of $D_\enm$ is referred to as the single ratio and is denoted as $S_\enm$. Both $D_\enm$ and $S_\enm$ benefit from cancellations of common experimental uncertainties and initial-state nuclear effects that are expected to modify all of the \PGUPnS states similarly~\cite{CMS:2026soy}.  The $D_\enm$ construction allows comparisons across collision systems of varying sizes without the need to account for the variation in the number of binary nucleon-nucleon collisions~\cite{CMS:2017ycw}. Final-state dissociation effects are expected to result in $D_\enm$ values below unity.  The \OO\ measurements are further reported as functions of charged-particle multiplicity and \PgU meson \pt. Tabulated results for this analysis can be found in the HEPData database~\cite{hepdata}.

The CMS apparatus~\cite{CMS:2008xjf,CMS:2023gfb} is a multipurpose, nearly hermetic detector designed to trigger on~\cite{CMS:2020cmk,CMS:2016ngn} and identify electrons, muons, photons, and hadrons~\cite{CMS:2015xaf,CMS:2018rym,CMS:2015myp,CMS:2014pgm}. 
A superconducting solenoid of 6\unit{m} internal diameter provides a magnetic field of 3.8\unit{T}. 
Within the solenoid volume are the silicon pixel and strip tracker, a crystal electromagnetic calorimeter, and a brass and scintillator hadron calorimeter. 
The pseudorapidity ($\eta$) coverage of these calorimeters is extended by the forward hadron  calorimeters, located at $3.0 < \abs{\eta} < 5.2$. 
Muons are measured in the range $\abs{\eta} < 2.4$ using gas-ionization detectors embedded in the steel flux-return yoke outside the solenoid. 
Events are filtered using a two-tiered trigger system~\cite{CMS:2016ngn}: the first level (L1), composed of custom hardware processors, uses information from the calorimeters and muon detectors~\cite{CMS:2020cmk}, while the high-level trigger executes a software-based event reconstruction.

The \PgU states are identified through their prompt decay into two oppositely charged muons whose momenta are precisely measured using the tracking and muon identification systems.  Events containing at least one (in \AAa data) or two (in \pp data) muons within a single bunch crossing are initially selected at L1. 
Reconstructed events are required to contain at least one reconstructed primary vertex~\cite{CMS-TDR-15-02}, ensuring genuine collision events. 
To suppress background from beam-gas interactions, at least 25\% of the reconstructed tracks are required to satisfy the ``high-purity'' criteria~\cite{CMS:2014pgm}.
Candidate \PgU mesons are formed offline from pairs of  muons identified using a particle-flow algorithm~\cite{CMS:2017yfk}.  Muon tracks are first reconstructed in the silicon tracker and then combined with compatible segments in the innermost muon chambers to ensure high purity, as most hadrons are absorbed by the calorimeters and support structures before reaching the muon system. 
Only muons within the kinematic acceptance $\pt>4\GeV$ and $\abs{\eta}<2.4$ are selected, ensuring well-understood reconstruction efficiency and acceptance. 
Muons are further required to satisfy the ``soft'' identification criteria~\cite{CMS:2024qgs}, demanding precise tracking and primary vertex association.
Muon pairs are required to have a common vertex.

{\tolerance=1200
Simulated Monte Carlo (MC) samples are generated separately for each \PgU state using \PYTHIA~8.311~\cite{Bierlich:2022pfr} with the CP5 tune~\cite{CMS:2019csb}, assuming unpolarized production~\cite{CMS:UpsPolarization,CMS:UpsPolarizationMult,LHCb:UpsPolarization,Acharya:2021PLB}. The generated \PgU mesons are weighted to reproduce the measured \pt spectra in the corresponding \OO, \NeNe, or \pp data~\cite{CMS:2020efs}. To model the underlying-event environment in ion collisions, the simulated \PgU events are embedded into minimum-bias \OO or \NeNe events generated with \HIJING~\cite{Wang:1991hta}. A full CMS detector simulation using \GEANTfour~\cite{Agostinelli:2002hh} is then used to derive acceptance and reconstruction efficiencies. Acceptance is defined as the fiducial phase space region of \PgU meson $\pt<30\GeV$ and rapidity $\abs{y}<2.4$. It is evaluated in simulations by calculating the fraction of \PgU candidates falling within that region and having decay muons passing the aforementioned muon kinematic selections. The reconstruction efficiency is defined as the fraction of accepted dimuon decays that are reconstructed and satisfy the trigger and offline selection requirements. Residual differences between data and simulation are corrected with single-muon scale factors derived from \PJGy\ meson tag-and-probe studies~\cite{CMS:2024qgs}.
\par}

\begin{figure*}[th]
  \centering
  \includegraphics[width=0.98\textwidth]{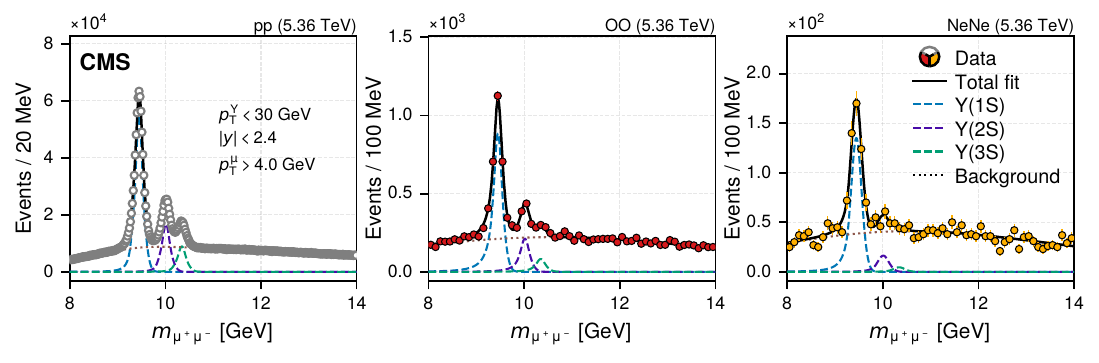}
  \caption{Muon pair invariant mass distributions in \pp\ (left), \OO\ (middle), and \NeNe\ (right) collisions. The overlaid curves show the corresponding fit results, integrated over muon pair $\pt<30\GeV$ and $\abs{y}<2.4$. In each case the total fit model is shown as a solid line, while the individual \PgU signal components and the background are shown as dashed lines.}
  \label{fig:fits}
\end{figure*}

The invariant mass of the muon pairs provides a distinct signature of the parent \PgU state and allows the three resonances to be separated. The single ratios are calculated from yields extracted with extended unbinned maximum likelihood fits to the dimuon invariant mass spectra, which are subsequently corrected for the acceptance and efficiency. As shown in Fig.~\ref{fig:fits}, the fits are performed in the range 8--14\GeV, which extends well beyond the invariant mass resolution around all three \PgU states. 
The total probability density function consists of three signal peaks, each described by the sum of two Crystal Ball functions~\cite{Oreglia:1980cs,Gaiser:1982yw} to account for the $\eta$-dependent detector resolution, and a background component modeled as an error function multiplied by an exponential. The peak positions and mass-resolution parameters of the \PgUb and \PgUc\ components are constrained relative to those of the \PgUa using the corresponding world-average mass ratios~\cite{ParticleDataGroup:2024cfk}.
The \pt-differential ratios are obtained by repeating the measurements in the intervals 0--3, 3--12, and 12--30\GeV.

 Systematic uncertainties are evaluated by varying either the analysis procedure or propagating external uncertainties. The signal and background model uncertainties are estimated by replacing the nominal signal and background shapes with alternative parameterizations that include an extra Crystal Ball function and a fourth-order Chebyshev polynomial, respectively. The uncertainty associated with the MC \pt weighting is estimated by varying the input \pt distributions within their uncertainties, and the uncertainty from the tag-and-probe correction is propagated from the corresponding single-muon scale factors. 

For $D_\enm$, the corresponding uncertainties in the \AAa and \pp samples are propagated to the final ratios. In \OO collisions, considering both the integrated and \pt-differential measurements, the total systematic uncertainties are in the ranges 1--6\% for \DBTwoOne, 6--16\% for \DBThreeOne, and 6--11\% for \DBThreeTwo. The lower ends of these ranges are generally associated with the integrated results. In the \pt-differential measurements, the background modeling is the dominant source of uncertainty, reflecting the lower statistical precision compared to the integrated results. Its contribution reaches about 5\% for \DBTwoOne, 15--16\% for \DBThreeOne, and 10--11\% for \DBThreeTwo.
The \pt\ weighting uncertainty is below about 1\% for most \DBTwoOne measurements, but reaches 5--6\% for ratios involving the \PgUc, especially in the integrated result. In \NeNe collisions, the integrated \DBTwoOne measurement has a total systematic uncertainty of about 3\%, driven by comparable signal and background model contributions. For $S_\enm$, reported as a function of charged-particle multiplicity in \OO collisions, the total experimental systematic uncertainty is in the ranges  0.1--6\% for \SNTwoOne\ and  6--16\% for \SNThreeOne. The uncertainty in \SNTwoOne\ is dominated by the background modeling, except for the lowest multiplicity interval, where smaller contributions from the signal model and \pt weighting are also relevant. For \SNThreeOne, the \pt weighting contributes about 6\% at low multiplicity, while the background model dominates at higher multiplicity, reaching about 15\%. The signal model uncertainty is generally below the percent level, and the tag-and-probe uncertainty is negligible in all multiplicity intervals. The total systematic uncertainty is computed as the quadratic sum of the individual contributions.

The measured double ratios in \OO collisions are
$\DBTwoOne=0.664 \pm 0.055\stat \pm 0.007\syst$,
$\DBThreeOne=0.392 \pm 0.077\stat \pm 0.023\syst$, and
$\DBThreeTwo=0.59 \pm 0.12\stat \pm 0.04\syst$.
In \NeNe collisions, $\DBTwoOne=0.27 \pm 0.12\stat \pm 0.01\syst$.
All double ratios are below unity, with significances above five standard deviations for \DBTwoOne and \DBThreeOne. The value of \DBThreeTwo in \OO collisions is also below unity by 3.2 standard deviations, providing the first evidence for sequential \PgU suppression in light ion collisions.

The \NeNe \DBTwoOne result provides a complementary check of system dependent effects in light ion collisions, as do the charged particle and open charm production studies recently reported by CMS~\cite{CMS:2026qef} and LHCb~\cite{LHCb:2026gwz}, respectively. The measured ratio is lower than the corresponding value in \OO by 3.0 standard deviations. 

\begin{figure}[ht!]
  \centering
  \includegraphics[width=0.48\textwidth]{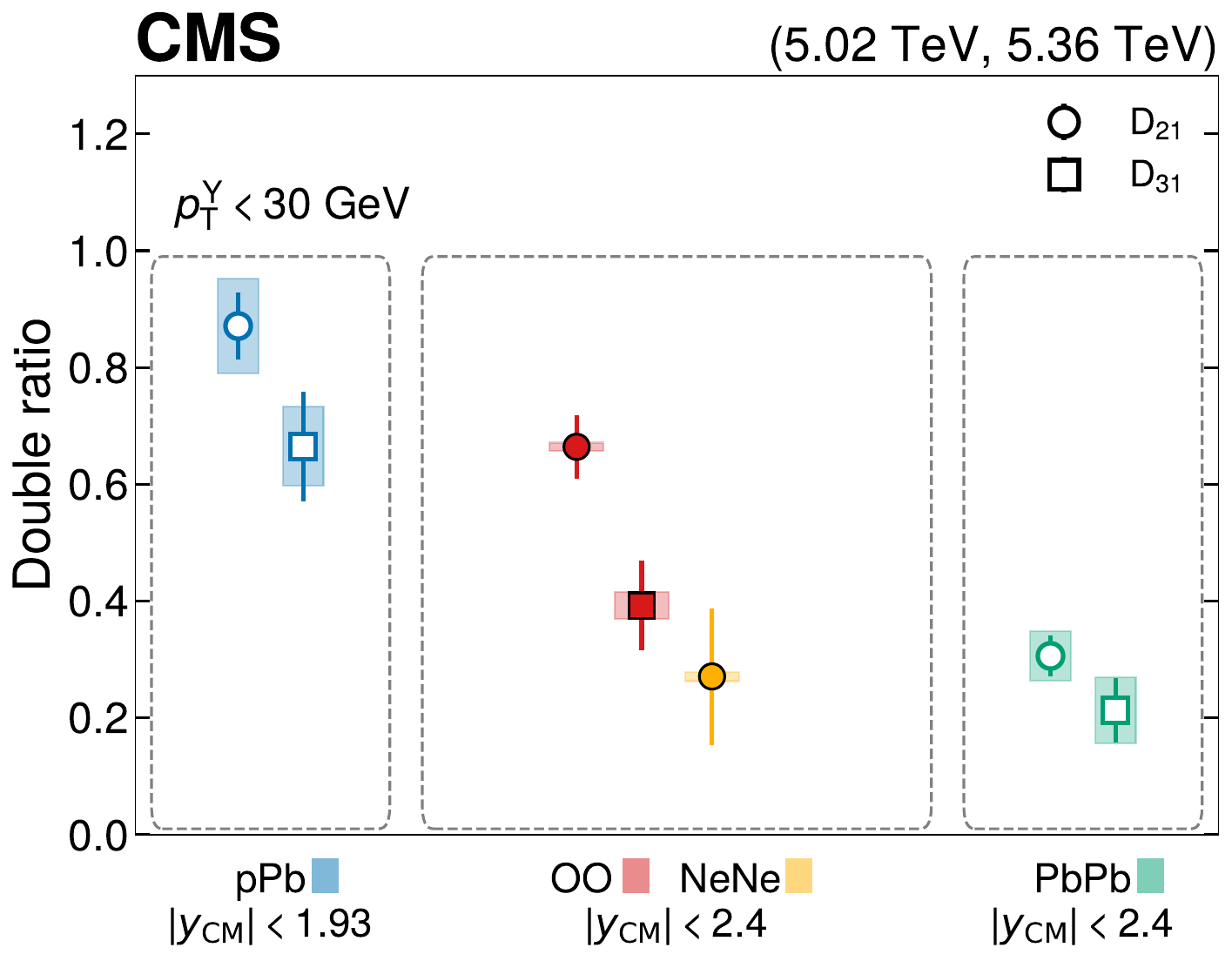}
  \includegraphics[width=0.48\textwidth]{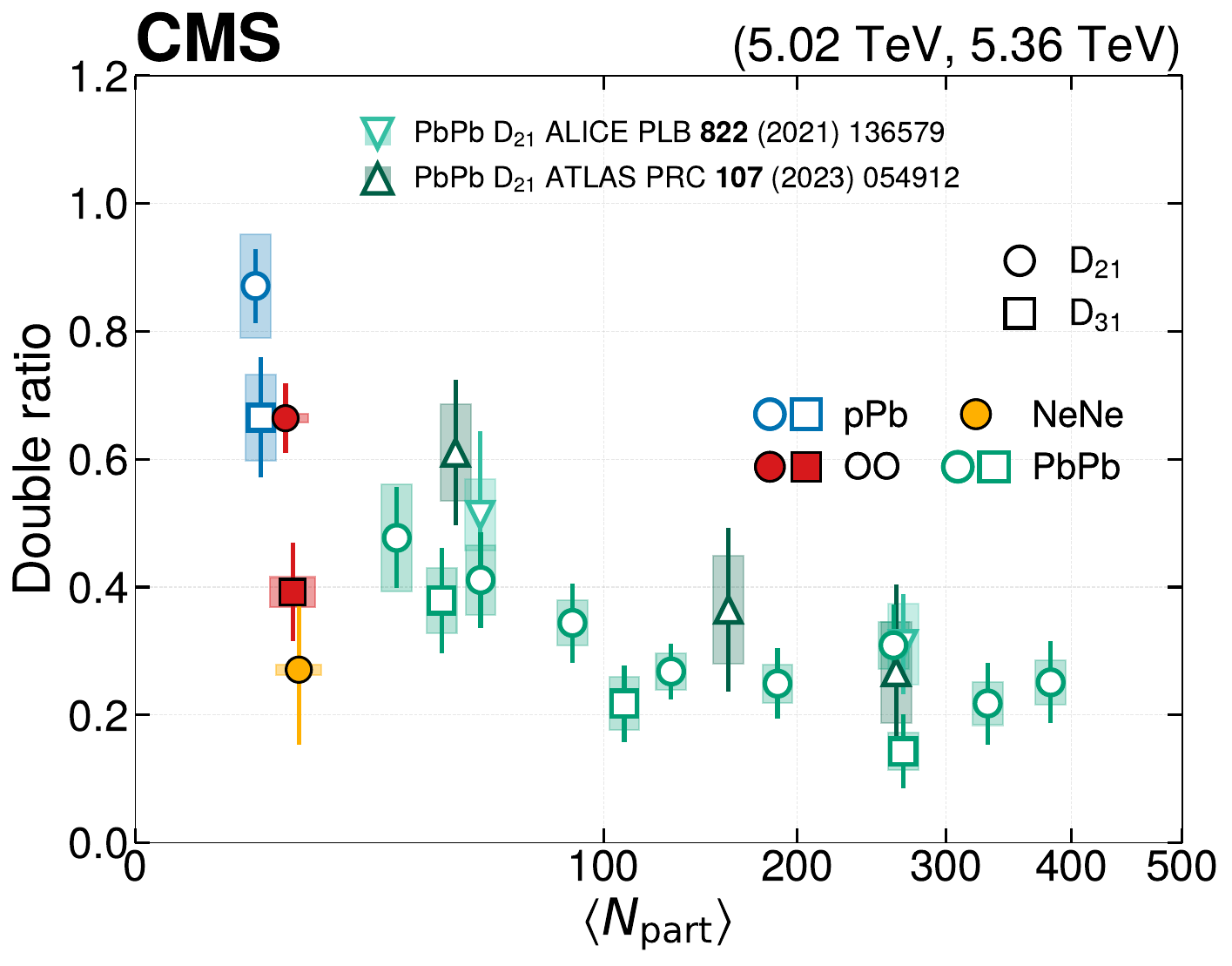}
  \caption{The measured double ratios \DBTwoOne\ and \DBThreeOne\ for \pPb~\cite{CMS:2022wfi}, \OO, \NeNe, and \PbPb~\cite{CMS:2018zza,CMS:2023lfu} collisions (\cmsLeft), exhibiting a clear hierarchy that evolves with increasing system size. The rapidity $y_{\mathrm{CM}}$ is defined in the center-of-mass (CM) frame.
  The double ratios are separately shown as a function of \ANpart (\cmsRight), and are compared with previous \PbPb~\cite{ALICE:2020wwx,ATLAS:2022exb} measurements performed at the LHC. For visibility purposes, the horizontal axis is displayed on a square root scale.}
  \label{fig:doubleratio}
\end{figure}

The left panel of Fig.~\ref{fig:doubleratio} summarizes  the measured \DBTwoOne and \DBThreeOne in \OO and \NeNe collisions, together with previous \pPb~\cite{CMS:2022wfi} and \PbPb~\cite{CMS:2018zza,CMS:2023lfu}  results, showing a clear evolution with the system size. For previous results not available directly in the ratio form shown here, common normalization uncertainties are removed where explicitly available; otherwise the uncertainties are propagated in $D_\enm$ or $S_\enm$ without assuming correlations and should be considered conservative. 

To place these measurements in context, the ratios are further studied as a function of system size. The average number of participating nucleons, $\langle \Npart \rangle$, is estimated with an MC Glauber model~\cite{Loizides:2025ule}, based on Woods--Saxon nuclear density distributions and the inelastic nucleon-nucleon cross section at the corresponding \sqrtsNN. For \pPb, \OO, and \NeNe collisions, the quoted $\langle \Npart \rangle$ values are obtained by averaging over minimum-bias events, whereas for \PbPb collisions they are averaged over the centrality intervals (\ie, the degree of overlap of the two ions) used in the corresponding measurements. This provides a geometry-based system size measure independent of final-state particle production.

The right panel of Fig.~\ref{fig:doubleratio} shows  \DBTwoOne and \DBThreeOne as a function of  \ANpart, together with \pPb~\cite{CMS:2022wfi} and \PbPb~\cite{ALICE:2020wwx,ATLAS:2022exb,CMS:2018zza,CMS:2023lfu} measurements performed at the LHC. The \OO and \NeNe results extend the system size dependence of excited-state \PgU suppression to $\ANpart \approx 10$, a region not previously accessed with symmetric ion collisions. The excited state suppression remains significant even at this smaller system size. The persistence of state-dependent \PgU suppression in \OO and \NeNe collisions indicates that final-state effects, \eg, those associated with deconfinement, can emerge in systems whose lifetime~\cite{Ke:2022gkq} and transverse area are smaller than in collisions of heavier ions. 

The corrected track multiplicity, \Ntrkcor, is used as an additional experimental handle for event activity. For all systems~\cite{CMS:2024qgs}, it is constructed from high-purity tracks with $\pt>0.4\GeV$ and $\abs{\eta}<2.4$ that are associated with the primary vertex. Event-by-event corrections account for detector acceptance, tracking efficiency, and residual misreconstruction effects, yielding a multiplicity estimator comparable across collision systems. Only events with a single reconstructed primary collision vertex are used, avoiding biases in the \Ntrkcor distribution from multiple collisions in a recorded event. The highest and lowest \Ntrkcor bins in \OO collisions correspond approximately to the 0--10\% and 40--100\% centrality intervals, respectively, with 0\% denoting the highest activity.

\begin{figure}[th!]
  \centering
  \includegraphics[width=0.48\textwidth]{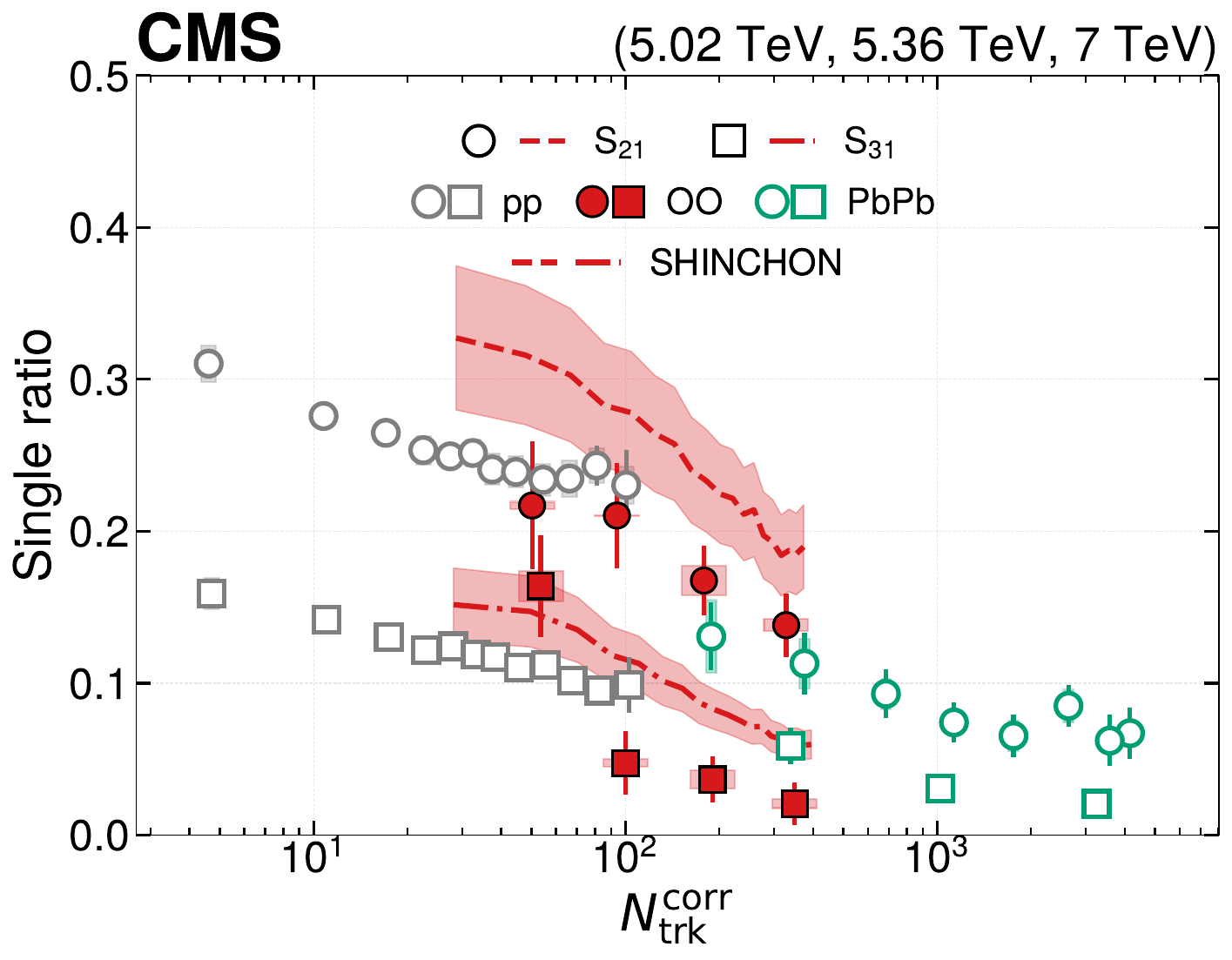}
  \includegraphics[width=0.48\textwidth]{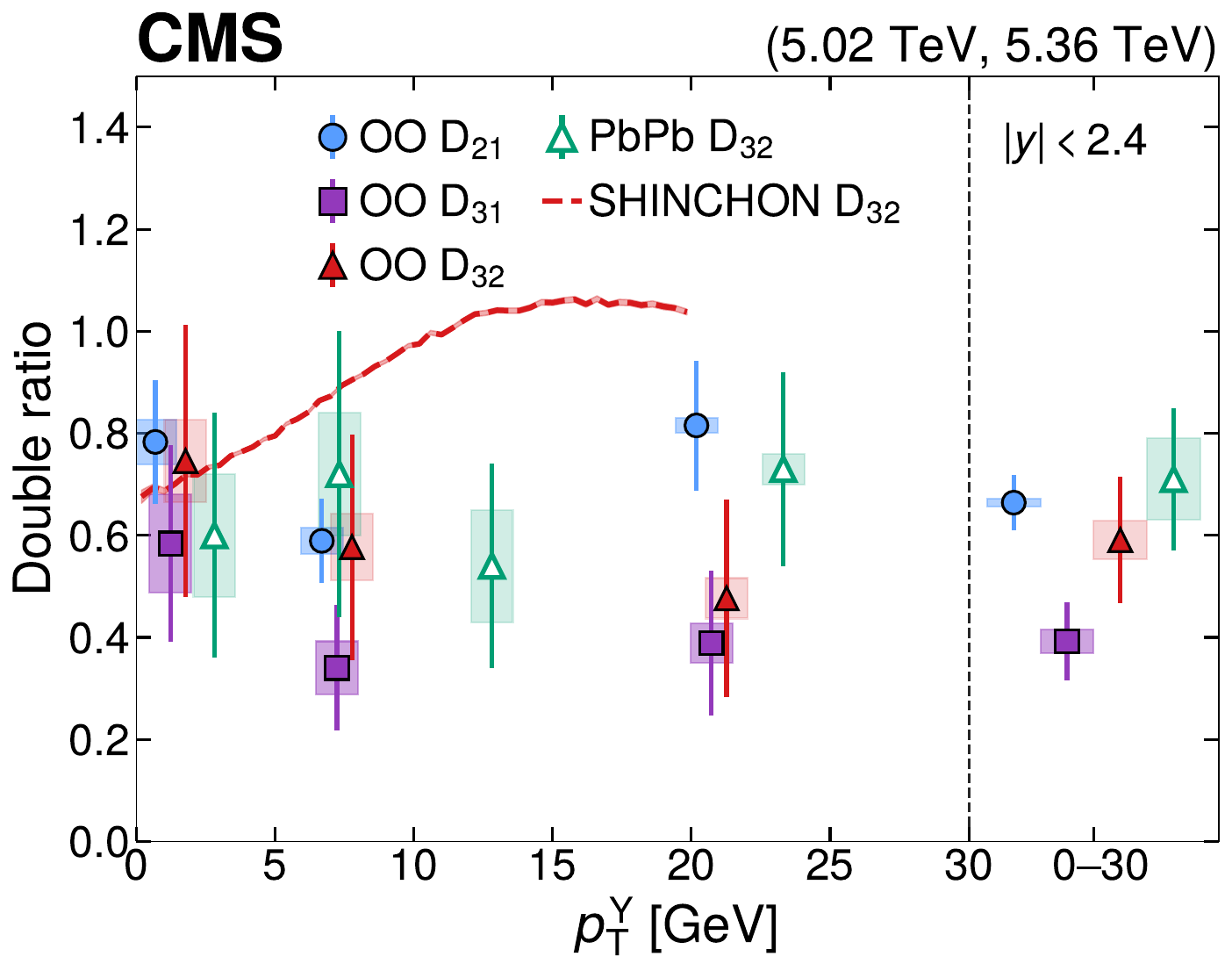}
  \caption{The single ratios \SNTwoOne and \SNThreeOne as a function of \Ntrkcor (\cmsLeft), overlaid with LHC measurements in \pp~\cite{CMS:2020fae} and \PbPb~\cite{CMS:2018zza,CMS:2023lfu} collisions. Measured double ratios \DBThreeTwo\ (\cmsRight) as a function of the muon pair \pt in \OO and \PbPb~\cite{CMS:2018zza,CMS:2023lfu} collisions, with the two points at the far right showing the \pt-integrated values for $\pt<30\GeV$ and $\abs{y}<2.4$.  For \OO collisions, predictions obtained with the \SHINCHON framework~\cite{Kim:2022lgu} are compared with the measured \SNTwoOne, \SNThreeOne, and \DBThreeTwo values. The shaded bands around the model predictions reflect the uncertainties in the calculation.}
  \label{fig:schematic}
\end{figure}

The left panel of Fig.~\ref{fig:schematic} shows the single ratios \SNTwoOne and \SNThreeOne as a function of \Ntrkcor. The \OO measurements extend the trend observed from \pp~\cite{CMS:2020fae} to \PbPb~\cite{CMS:2018zza,CMS:2023lfu} collisions, indicating a clear evolution of quarkonium suppression with increasing event activity. While the present uncertainties and possible differences in feed-down contributions preclude assigning a standalone significance to a comparison between collision systems, the trend should be carefully interpreted because event activity alone does not fully determine the observed suppression. Additional sensitivity to the collision geometry, initial nuclear environment, or medium lifetime in symmetric light ion collisions should be considered. Indeed, 
measurements of charged-hadron production show that head-on \OO collisions can achieve particle production per participant comparable to the head-on \PbPb case at similar \sqrtsNN~\cite{ATLAS:2026zgq,CMS:2026sai}, indicating that the smaller \OO event activity may not imply a dilute medium.

The right panel of Fig.~\ref{fig:schematic} presents the \pt-differential measurements of \DBTwoOne, \DBThreeOne, and \DBThreeTwo, along with their \pt-integrated values.  Within uncertainties, no particular \pt dependence is seen. The \DBThreeTwo values measured in \OO and \PbPb collisions are compared and found to be compatible within uncertainties across the measured \pt range and in the integrated region. This comparison in different kinematic ranges complements the study of system size dependence and supports a common interpretation of sequential \PgU suppression in \OO and \PbPb collisions, with the additional depletion of \PgUc relative to \PgUb providing particular sensitivity to the hottest stages of the medium.

Figure~\ref{fig:schematic} also compares the measured \SNTwoOne, \SNThreeOne, and \DBThreeTwo with predictions from the \SHINCHON framework~\cite{Kim:2022lgu}. This framework combines an event-by-event hydrodynamic description of the medium with a bottomonium dissociation calculation based on gluo-dissociation and inelastic parton scattering. Feed-down contributions to the inclusive \PgU yields are included, while regeneration is neglected in the \OO implementation. The model predictions include the statistical precision in the calculation and the uncertainty in the \pp baseline ratios used to set the absolute normalization. This \pp normalization uncertainty cancels in the $D_\enm$ ratio. The comparison provides a useful benchmark for interpreting the observed sequential suppression in terms of final-state medium effects. The current calculation significantly underestimates the measured suppression, suggesting that additional medium dynamics or stronger dissociation effects may be required to describe the data.  This tension is especially relevant for the \pt-differential comparison, where the assumed \PGUPnS formation time can have a sizable impact at high \pt, while its effect on the \Ntrkcor dependence is expected to be subleading because most reconstructed \PgU mesons have \pt around 4--5\GeV. These results help establish a coherent picture of bottomonium behavior across a range of collision system sizes. They provide essential input for determining the underlying production mechanisms across \pt and whether in-medium \PgU meson suppression emerges gradually with increasing event activity, rather than through an abrupt threshold behavior.

In summary, the \PgUa, \PgUb, and \PgUc states are measured for the first time in oxygen-oxygen collisions, with the \PgUa and \PgUb states also measured in neon-neon collisions. The data were recorded with the CMS detector at the LHC in 2025 and collected at a nucleon-nucleon center-of-mass energy of 5.36\TeV. Proton-proton (\pp) collisions at the same energy provided the reference baseline. The suppression of these states is quantified through ratios that compare excited-to-reference-state \PgU meson yields within ion-ion collisions and, separately, relative to the corresponding \pp yields. These observables isolate final-state medium effects by canceling many common experimental uncertainties and reducing sensitivity to initial-state effects.

The measurements reveal significant suppression of the \PgUb and \PgUc states, with significances exceeding five standard deviations when measured relative to \PgUa. The suppression of \PgUc relative to \PgUb exceeds three standard deviations, providing evidence for sequential \PGUPnS meson suppression in light ion collisions. Moreover, the overall magnitude and transverse momentum dependence of the \PgUc{}/\PgUb ratio are compatible between oxygen-oxygen and lead-lead collisions, supporting a common interpretation of sequential \PgU meson suppression in the two systems.  These results reinforce earlier findings that quarkonium suppression may not require large volumes of matter and provide new constraints for theoretical models aiming to describe bottomonium production from \pp to heavy ion collisions.

\begin{acknowledgments}
We congratulate our colleagues in the CERN accelerator departments for the excellent performance of the LHC and thank the technical and administrative staffs at CERN and at other CMS institutes for their contributions to the success of the CMS effort. In addition, we gratefully acknowledge the computing centers and personnel of the Worldwide LHC Computing Grid and other centers for delivering so effectively the computing infrastructure essential to our analyses. Finally, we acknowledge the enduring support for the construction and operation of the LHC, the CMS detector, and the supporting computing infrastructure provided by the following funding agencies: SC (Armenia), BMFWF and FWF (Austria); FNRS and FWO (Belgium); CNPq, CAPES, FAPERJ, FAPERGS, and FAPESP (Brazil); MES and BNSF (Bulgaria); CERN; CAS, MoST, and NSFC (China); MINCIENCIAS (Colombia); MSES and CSF (Croatia); RIF (Cyprus); SENESCYT (Ecuador); ERC PRG and PSG, TARISTU24-TK10 and MoER TK202 (Estonia); Academy of Finland, MEC, and HIP (Finland); CEA and CNRS/IN2P3 (France); SRNSF (Georgia); BMFTR, DFG, and HGF (Germany); GSRI (Greece); MATE and NKFIH (Hungary); DAE and DST (India); IPM (Iran); SFI (Ireland); INFN (Italy); MSIT and NRF (Republic of Korea); MES (Latvia); LMTLT (Lithuania); MOE and UM (Malaysia); BUAP, CINVESTAV, CONACYT, LNS, SEP, and UASLP-FAI (Mexico); MOS (Montenegro); MBIE (New Zealand); PAEC (Pakistan); MSHE, NSC, and NAWA (Poland); FCT (Portugal); MESTD (Serbia); MICIU/AEI and PCTI (Spain); MOSTR (Sri Lanka); Swiss Funding Agencies (Switzerland); MST (Taipei); MHESI (Thailand); TUBITAK and TENMAK (T\"{u}rkiye); NASU (Ukraine); STFC (United Kingdom); DOE and NSF (USA).
\end{acknowledgments}\section*{Data availability} Release and preservation of data used by the CMS Collaboration as the basis for publications is guided by the  \href{https://doi.org/10.7483/OPENDATA.CMS.1BNU.8V1W}{CMS data preservation, re-use and open access policy}.

\bibliography{auto_generated}
\cleardoublepage \appendix\section{The CMS Collaboration \label{app:collab}}\begin{sloppypar}\hyphenpenalty=5000\widowpenalty=500\clubpenalty=5000\input{HIN-25-015-public-authorlist.tex}\end{sloppypar}
%%% END EDITABLE REGION %%%
% skeleton_end
\end{document}

%% file: HIN-25-015-public-authorlist.tex
\cmsinstitute{Yerevan Physics Institute, Yerevan, Armenia}
{\tolerance=6000
A.~Belyaev\cmsorcid{0000-0003-1692-1173}, A.~Gevorgyan\cmsorcid{0000-0003-2751-9489}, A.~Hayrapetyan, A.~Tumasyan\cmsAuthorMark{1}\cmsorcid{0009-0000-0684-6742}
\par}
\cmsinstitute{Institut f\"{u}r Hochenergiephysik, Vienna, Austria}
{\tolerance=6000
P.S.~Hussain\cmsorcid{0000-0002-4825-5278}, M.~Sonawane\cmsorcid{0000-0003-0510-7010}
\par}
\cmsinstitute{Marietta Blau Institute for Particle Physics, Vienna, Austria}
{\tolerance=6000
W.~Adam\cmsorcid{0000-0001-9099-4341}, L.~Benato\cmsorcid{0000-0001-5135-7489}, T.~Bergauer\cmsorcid{0000-0002-5786-0293}, M.~Dragicevic\cmsorcid{0000-0003-1967-6783}, S.~Gundacker\cmsorcid{0000-0003-2087-3266}, A.K.~Guven\cmsorcid{0009-0004-5670-5138}, M.~Jeitler\cmsAuthorMark{2}\cmsorcid{0000-0002-5141-9560}, N.~Krammer\cmsorcid{0000-0002-0548-0985}, A.~Li\cmsorcid{0000-0002-4547-116X}, D.~Liko\cmsorcid{0000-0002-3380-473X}, M.~Matthewman, A.~Pfeiffer\cmsorcid{0000-0001-5328-448X}, J.~Schieck\cmsAuthorMark{2}\cmsorcid{0000-0002-1058-8093}, R.~Sch\"{o}fbeck\cmsAuthorMark{2}\cmsorcid{0000-0002-2332-8784}, M.~Shooshtari\cmsorcid{0009-0004-8882-4887}, N.~Van~Den~Bossche\cmsorcid{0000-0003-2973-4991}, W.~Waltenberger\cmsorcid{0000-0002-6215-7228}, C.E.~Wulz\cmsAuthorMark{2}\cmsorcid{0000-0001-9226-5812}
% ...
\par}
$^{1}$Also at Yerevan State University, Yerevan, Armenia\\
$^{2}$Also at Technische Universit\"{a}t Wien, Vienna, Austria\\
% ...


%% file: auto_generated.bib
@ARTICLE{Ellis:2025nkm,
	AUTHOR=	"Ellis, John",
	TITLE=	"{Personal memories of 50 years of quarkonia}",
	EPRINT=	"2506.10643",
	ARCHIVEPREFIX=	"arXiv",
	PRIMARYCLASS=	"hep-ph",
	REPORTNUMBER=	"KCL-PH-TH/2025-22, CERN-TH-2025-115",
	DOI=	"10.1016/j.nuclphysb.2025.117062",
	JOURNAL=	"Nucl. Phys. B",
	VOLUME=	"1018",
	PAGES=	"117062",
	YEAR=	"2025",
}

@ARTICLE{Harris:1996zx,
	AUTHOR=	"Harris, John W. and Muller, Berndt",
	TITLE=	"{The search for the quark-gluon plasma}",
	EPRINT=	"hep-ph/9602235",
	ARCHIVEPREFIX=	"arXiv",
	REPORTNUMBER=	"DUKE-TH-96-105",
	DOI=	"10.1146/annurev.nucl.46.1.71",
	JOURNAL=	"Ann. Rev. Nucl. Part. Sci.",
	VOLUME=	"46",
	PAGES=	"71",
	YEAR=	"1996",
}

@ARTICLE{Matsui:1986dk,
	AUTHOR=	"Matsui, T. and Satz, H.",
	TITLE=	"{\PJGy suppression by quark-gluon plasma formation}",
	REPORTNUMBER=	"BNL-38344",
	DOI=	"10.1016/0370-2693(86)91404-8",
	JOURNAL=	"Phys. Lett. B",
	VOLUME=	"178",
	PAGES=	"416",
	YEAR=	"1986",
}

@ARTICLE{Andronic:2025jbp,
	AUTHOR=	"Andronic, Anton and Arnaldi, Roberta",
	TITLE=	"{Quarkonia and deconfined quark-gluon matter in heavy ion collisions}",
	EPRINT=	"2501.08290",
	ARCHIVEPREFIX=	"arXiv",
	PRIMARYCLASS=	"nucl-ex",
	DOI=	"10.1146/annurev-nucl-121423-101041",
	JOURNAL=	"Ann. Rev. Nucl. Part. Sci.",
	VOLUME=	"75",
	PAGES=	"351",
	YEAR=	"2025",
}

@ARTICLE{Apolinario:2022vzg,
	AUTHOR=	"Apolin{\'a}rio, Liliana and Lee, Yen-Jie and Winn, Michael",
	TITLE=	"{Heavy quarks and jets as probes of the {QGP}}",
	EPRINT=	"2203.16352",
	ARCHIVEPREFIX=	"arXiv",
	PRIMARYCLASS=	"hep-ph",
	DOI=	"10.1016/j.ppnp.2022.103990",
	JOURNAL=	"Prog. Part. Nucl. Phys.",
	VOLUME=	"127",
	PAGES=	"103990",
	YEAR=	"2022",
}

@ARTICLE{Thews:2000rj,
	AUTHOR=	"Thews, Robert L. and Schroedter, Martin and Rafelski, Johann",
	TITLE=	"{Enhanced \PJGy production in deconfined quark matter}",
	EPRINT=	"hep-ph/0007323",
	ARCHIVEPREFIX=	"arXiv",
	DOI=	"10.1103/PhysRevC.63.054905",
	JOURNAL=	"Phys. Rev. C",
	VOLUME=	"63",
	PAGES=	"054905",
	YEAR=	"2001",
}

@ARTICLE{Braun-Munzinger:2000csl,
	AUTHOR=	"Braun-Munzinger, P. and Stachel, J.",
	TITLE=	"{(Non)thermal aspects of charmonium production and a new look at \PJGy suppression}",
	EPRINT=	"nucl-th/0007059",
	ARCHIVEPREFIX=	"arXiv",
	DOI=	"10.1016/S0370-2693(00)00991-6",
	JOURNAL=	"Phys. Lett. B",
	VOLUME=	"490",
	PAGES=	"196",
	YEAR=	"2000",
}

@ARTICLE{Du:2017qkv,
	AUTHOR=	"Du, Xiaojian and He, Min and Rapp, Ralf",
	TITLE=	"{Color screening and regeneration of bottomonia in high-energy heavy ion collisions}",
	EPRINT=	"1706.08670",
	ARCHIVEPREFIX=	"arXiv",
	PRIMARYCLASS=	"hep-ph",
	DOI=	"10.1103/PhysRevC.96.054901",
	JOURNAL=	"Phys. Rev. C",
	VOLUME=	"96",
	PAGES=	"054901",
	YEAR=	"2017",
}

@ARTICLE{Yao:2020xzw,
	AUTHOR=	"Yao, Xiaojun and Ke, Weiyao and Xu, Yingru and Bass, Steffen A. and M{\"u}ller, Berndt",
	TITLE=	"{Coupled {Boltzmann} transport equations of heavy quarks and quarkonia in quark-gluon plasma}",
	EPRINT=	"2004.06746",
	ARCHIVEPREFIX=	"arXiv",
	PRIMARYCLASS=	"hep-ph",
	REPORTNUMBER=	"MIT-CTP/5192",
	DOI=	"10.1007/JHEP01(2021)046",
	JOURNAL=	"JHEP",
	VOLUME=	"01",
	PAGES=	"046",
	YEAR=	"2021",
}

@ARTICLE{Brambilla:2023hkw,
	AUTHOR=	"Brambilla, Nora and Escobedo, Miguel {\'A}ngel and Islam, Ajaharul and Strickland, Michael and Tiwari, Anurag and Vairo, Antonio and Vander Griend, Peter",
	TITLE=	"{Regeneration of bottomonia in an open quantum systems approach}",
	EPRINT=	"2302.11826",
	ARCHIVEPREFIX=	"arXiv",
	PRIMARYCLASS=	"hep-ph",
	REPORTNUMBER=	"TUM-EFT 178/23, FERMILAB-PUB-23-060-V",
	DOI=	"10.1103/PhysRevD.108.L011502",
	JOURNAL=	"Phys. Rev. D",
	VOLUME=	"108",
	PAGES=	"L011502",
	YEAR=	"2023",
}

@ARTICLE{Wu:2025lcj,
	AUTHOR=	"Wu, Biaogang and Rapp, Ralf",
	TITLE=	"{Bottomonium transport in a strongly coupled quark-gluon plasma}",
	EPRINT=	"2508.20995",
	ARCHIVEPREFIX=	"arXiv",
	PRIMARYCLASS=	"nucl-th",
	DOI=	"10.1016/j.physletb.2026.140223",
	JOURNAL=	"Phys. Lett. B",
	VOLUME=	"873",
	PAGES=	"140223",
	YEAR=	"2026",
}

@ARTICLE{Bahavar:2014jza,
	AUTHOR=	"Bahavar, Philipp B. and Uphoff, Jan and Greiner, Carsten",
	TITLE=	"{Hadronic versus partonic \PJGy production in the statistical hadronization model}",
	EPRINT=	"1404.4517",
	ARCHIVEPREFIX=	"arXiv",
	PRIMARYCLASS=	"hep-ph",
	DOI=	"10.1103/PhysRevC.90.061901",
	JOURNAL=	"Phys. Rev. C",
	VOLUME=	"90",
	PAGES=	"061901",
	YEAR=	"2014",
}

@ARTICLE{Andronic:2022ucg,
	AUTHOR=	"Andronic, A. and Braun-Munzinger, P. and Redlich, K. and Stachel, J.",
	TITLE=	"{Statistical hadronization of \PQb quarks in \PbPb collisions at {LHC} energy: a case for partial equilibration of \PQb quarks?}",
	EPRINT=	"2209.14562",
	ARCHIVEPREFIX=	"arXiv",
	PRIMARYCLASS=	"hep-ph",
	DOI=	"10.5506/APhysPolBSupp.16.1-A107",
	JOURNAL=	"Acta Phys. Polon. Supp.",
	VOLUME=	"16",
	PAGES=	"1",
	YEAR=	"2023",
}

@ARTICLE{Andronic:2024oxz,
	AUTHOR=	"Andronic, A. and others",
	TITLE=	"{Comparative study of quarkonium transport in hot {QCD} matter}",
	EPRINT=	"2402.04366",
	ARCHIVEPREFIX=	"arXiv",
	PRIMARYCLASS=	"nucl-th",
	REPORTNUMBER=	"FERMILAB-PUB-24-0005-T-V",
	DOI=	"10.1140/epja/s10050-024-01306-6",
	JOURNAL=	"Eur. Phys. J. A",
	VOLUME=	"60",
	PAGES=	"88",
	YEAR=	"2024",
}

@ARTICLE{ParticleDataGroup:2024cfk,
	AUTHOR=	"{Particle Data Group} and Navas, S. and others",
	TITLE=	"{Review of particle physics}",
	DOI=	"10.1103/PhysRevD.110.030001",
	JOURNAL=	"Phys. Rev. D",
	VOLUME=	"110",
	PAGES=	"030001",
	YEAR=	"2024",
}

@ARTICLE{CMS:2024krd,
	AUTHOR=	"{CMS Collaboration}",
	TITLE=	"{Overview of high-density QCD studies with the CMS experiment at the LHC}",
	EPRINT=	"2405.10785",
	ARCHIVEPREFIX=	"arXiv",
	PRIMARYCLASS=	"nucl-ex",
	REPORTNUMBER=	"CMS-HIN-23-011, CERN-EP-2024-057",
	DOI=	"10.1016/j.physrep.2024.11.007",
	JOURNAL=	"Phys. Rept.",
	VOLUME=	"1115",
	PAGES=	"219",
	YEAR=	"2025",
}

@ARTICLE{CMS:2011all,
	AUTHOR=	"{CMS Collaboration}",
	TITLE=	"{Indications of suppression of excited \PgU states in \PbPb collisions at $\sqrtsNN = 2.76\TeV$}",
	EPRINT=	"1105.4894",
	ARCHIVEPREFIX=	"arXiv",
	PRIMARYCLASS=	"nucl-ex",
	REPORTNUMBER=	"CERN-PH-EP-2011-074, CMS-HIN-11-007",
	DOI=	"10.1103/PhysRevLett.107.052302",
	JOURNAL=	"Phys. Rev. Lett.",
	VOLUME=	"107",
	PAGES=	"052302",
	YEAR=	"2011",
}

@ARTICLE{CMS:2012gvv,
	AUTHOR=	"{CMS Collaboration}",
	TITLE=	"{Observation of sequential \PgU suppression in \PbPb collisions}",
	EPRINT=	"1208.2826",
	ARCHIVEPREFIX=	"arXiv",
	PRIMARYCLASS=	"nucl-ex",
	REPORTNUMBER=	"CMS-HIN-11-011, CERN-PH-EP-2012-228",
	DOI=	"10.1103/PhysRevLett.109.222301",
	JOURNAL=	"Phys. Rev. Lett.",
	VOLUME=	"109",
	PAGES=	"222301",
	YEAR=	"2012",
	NOTE=	"[Erratum: \DOI{10.1103/PhysRevLett.120.199903}]",
}

@ARTICLE{CMS:2018zza,
	AUTHOR=	"{CMS Collaboration}",
	TITLE=	"{Measurement of nuclear modification factors of \PgUa, \PgUb, and \PgUc mesons in \PbPb collisions at $\sqrtsNN = 5.02\TeV$}",
	EPRINT=	"1805.09215",
	ARCHIVEPREFIX=	"arXiv",
	PRIMARYCLASS=	"hep-ex",
	REPORTNUMBER=	"CMS-HIN-16-023, CERN-EP-2018-110",
	DOI=	"10.1016/j.physletb.2019.01.006",
	JOURNAL=	"Phys. Lett. B",
	VOLUME=	"790",
	PAGES=	"270",
	YEAR=	"2019",
}

@ARTICLE{CMS:2023lfu,
	AUTHOR=	"{CMS Collaboration}",
	TITLE=	"{Observation of the \PgUc meson and suppression of \PgU states in \PbPb collisions at $\sqrtsNN = 5.02\TeV$}",
	EPRINT=	"2303.17026",
	ARCHIVEPREFIX=	"arXiv",
	PRIMARYCLASS=	"hep-ex",
	REPORTNUMBER=	"CMS-HIN-21-007, CERN-EP-2023-011",
	DOI=	"10.1103/PhysRevLett.133.022302",
	JOURNAL=	"Phys. Rev. Lett.",
	VOLUME=	"133",
	PAGES=	"022302",
	YEAR=	"2024",
}

@ARTICLE{ALICE:2020wwx,
	AUTHOR=	"{ALICE Collaboration}",
	TITLE=	"{\PgU production and nuclear modification at forward rapidity in \PbPb collisions at $\sqrtsNN = 5.02\TeV$}",
	EPRINT=	"2011.05758",
	ARCHIVEPREFIX=	"arXiv",
	PRIMARYCLASS=	"nucl-ex",
	REPORTNUMBER=	"CERN-EP-2020-207",
	DOI=	"10.1016/j.physletb.2021.136579",
	JOURNAL=	"Phys. Lett. B",
	VOLUME=	"822",
	PAGES=	"136579",
	YEAR=	"2021",
}

@ARTICLE{ATLAS:2022exb,
	AUTHOR=	"{ATLAS Collaboration}",
	TITLE=	"{Production of \PgU(n\text{S}) mesons in \PbPb and \pp collisions at $5.02\TeV$}",
	EPRINT=	"2205.03042",
	ARCHIVEPREFIX=	"arXiv",
	PRIMARYCLASS=	"nucl-ex",
	REPORTNUMBER=	"CERN-EP-2022-045",
	DOI=	"10.1103/PhysRevC.107.054912",
	JOURNAL=	"Phys. Rev. C",
	VOLUME=	"107",
	PAGES=	"054912",
	YEAR=	"2023",
}

@ARTICLE{STAR:2022rpk,
	AUTHOR=	"Aboona, Bassam and others",
	COLLABORATION=	"STAR",
	TITLE=	"{Observation of sequential \PgU suppression in {AuAu} collisions at $\sqrtsNN = 200\GeV$ with the {STAR} experiment}",
	EPRINT=	"2207.06568",
	ARCHIVEPREFIX=	"arXiv",
	PRIMARYCLASS=	"nucl-ex",
	DOI=	"10.1103/PhysRevLett.130.112301",
	JOURNAL=	"Phys. Rev. Lett.",
	VOLUME=	"130",
	PAGES=	"112301",
	YEAR=	"2023",
}

@ARTICLE{ALICE:2019qie,
	AUTHOR=	"{ALICE Collaboration}",
	TITLE=	"{\PgU production in \pPb collisions at $\sqrtsNN = 8.16\TeV$}",
	EPRINT=	"1910.14405",
	ARCHIVEPREFIX=	"arXiv",
	PRIMARYCLASS=	"nucl-ex",
	REPORTNUMBER=	"CERN-EP-2019-243",
	DOI=	"10.1016/j.physletb.2020.135486",
	JOURNAL=	"Phys. Lett. B",
	VOLUME=	"806",
	PAGES=	"135486",
	YEAR=	"2020",
}

@ARTICLE{ATLAS:2017prf,
	AUTHOR=	"{ATLAS Collaboration}",
	TITLE=	"{Measurement of quarkonium production in \pPb and \pp collisions at $5.02\TeV$ with the {ATLAS} detector}",
	EPRINT=	"1709.03089",
	ARCHIVEPREFIX=	"arXiv",
	PRIMARYCLASS=	"nucl-ex",
	REPORTNUMBER=	"CERN-EP-2017-197, CERN-PH-EP-2017-197",
	DOI=	"10.1140/epjc/s10052-018-5624-4",
	JOURNAL=	"Eur. Phys. J. C",
	VOLUME=	"78",
	PAGES=	"171",
	YEAR=	"2018",
}

@ARTICLE{CMS:2022wfi,
	AUTHOR=	"{CMS Collaboration}",
	TITLE=	"{Nuclear modification of \PgU states in \pPb collisions at $\sqrtsNN = 5.02\TeV$}",
	EPRINT=	"2202.11807",
	ARCHIVEPREFIX=	"arXiv",
	PRIMARYCLASS=	"hep-ex",
	REPORTNUMBER=	"CMS-HIN-18-005, CERN-EP-2020-181",
	DOI=	"10.1016/j.physletb.2022.137397",
	JOURNAL=	"Phys. Lett. B",
	VOLUME=	"835",
	PAGES=	"137397",
	YEAR=	"2022",
}

@ARTICLE{LHCb:2018psc,
	AUTHOR=	"{LHCb Collaboration}",
	TITLE=	"{Study of \PgU production in \pPb collisions at $\sqrtsNN = 8.16\TeV$}",
	EPRINT=	"1810.07655",
	ARCHIVEPREFIX=	"arXiv",
	PRIMARYCLASS=	"hep-ex",
	REPORTNUMBER=	"LHCb-PAPER-2018-035, CERN-EP-2018-267",
	DOI=	"10.1007/JHEP11(2018)194",
	JOURNAL=	"JHEP",
	VOLUME=	"11",
	PAGES=	"194",
	YEAR=	"2018",
	NOTE=	"[Erratum: \DOI{10.1007/JHEP02(2020)093}]",
}

@ARTICLE{Emerick:2011xu,
	AUTHOR=	"Emerick, A. and Zhao, X. and Rapp, R.",
	TITLE=	"{Bottomonia in the quark gluon plasma and their production at {RHIC} and {LHC}}",
	JOURNAL=	"Eur. Phys. J. A",
	VOLUME=	"48",
	YEAR=	"2012",
	PAGES=	"72",
	DOI=	"10.1140/epja/i2012-12072-y",
	EPRINT=	"1111.6537",
	ARCHIVEPREFIX=	"arXiv",
	PRIMARYCLASS=	"hep-ph",
}

@ARTICLE{Bagchi:2023vfv,
	AUTHOR=	"Bagchi, Partha and Das, Arpan and Mishra, Ananta P.",
	TITLE=	"{Does quarkonia suppression serve as a probe for the deconfinement in small systems?}",
	EPRINT=	"2310.12267",
	ARCHIVEPREFIX=	"arXiv",
	PRIMARYCLASS=	"nucl-th",
	DOI=	"10.1103/PhysRevD.110.014017",
	JOURNAL=	"Phys. Rev. D",
	VOLUME=	"110",
	PAGES=	"014017",
	YEAR=	"2024",
}

@ARTICLE{Sahoo:2019ifs,
	AUTHOR=	"Sahoo, Raghunath",
	TITLE=	"{Possible formation of {QGP}-droplets in proton-proton collisions at the {CERN} {Large Hadron Collider}}",
	EPRINT=	"1908.10566",
	ARCHIVEPREFIX=	"arXiv",
	PRIMARYCLASS=	"nucl-ex",
	DOI=	"10.22661/AAPPSBL.2019.29.4.16",
	JOURNAL=	"AAPPS Bull.",
	VOLUME=	"29",
	PAGES=	"16",
	YEAR=	"2019",
}

@ARTICLE{Gardim:2022yds,
	AUTHOR=	"Gardim, Fernando G. and Krupczak, Renata and da Silva, Tiago Nunes",
	TITLE=	"{Smallest drop of {QGP}: thermodynamic properties of \pPb collisions}",
	EPRINT=	"2212.11710",
	ARCHIVEPREFIX=	"arXiv",
	PRIMARYCLASS=	"nucl-th",
	DOI=	"10.1103/PhysRevC.109.014904",
	JOURNAL=	"Phys. Rev. C",
	VOLUME=	"109",
	PAGES=	"014904",
	YEAR=	"2024",
}

@ARTICLE{Ke:2022gkq,
	AUTHOR=	"Ke, Weiyao and Vitev, Ivan",
	TITLE=	"{Searching for {QGP} droplets with high-$\pt$ hadrons and heavy flavor}",
	EPRINT=	"2204.00634",
	ARCHIVEPREFIX=	"arXiv",
	PRIMARYCLASS=	"hep-ph",
	REPORTNUMBER=	"LA-UR-22-23062",
	DOI=	"10.1103/PhysRevC.107.064903",
	JOURNAL=	"Phys. Rev. C",
	VOLUME=	"107",
	PAGES=	"064903",
	YEAR=	"2023",
}

@ARTICLE{PHENIX:2023dxl,
	AUTHOR=	"Abdulameer, N. J. and others",
	COLLABORATION=	"PHENIX",
	TITLE=	"{Disentangling centrality bias and final-state effects in the production of high-$\pt$ neutral pions using direct photon in d+{Au} collisions at $\sqrtsNN = 200\GeV$}",
	EPRINT=	"2303.12899",
	ARCHIVEPREFIX=	"arXiv",
	PRIMARYCLASS=	"nucl-ex",
	DOI=	"10.1103/PhysRevLett.134.022302",
	JOURNAL=	"Phys. Rev. Lett.",
	VOLUME=	"134",
	PAGES=	"022302",
	YEAR=	"2025",
}

@ARTICLE{FerreiroComp,
	AUTHOR=	"Ferreiro, Elena G. and Lansberg, Jean-Philippe",
	TITLE=	"{Is bottomonium suppression in proton-nucleus and nucleus-nucleus collisions at {LHC} energies due to the same effects?}",
	JOURNAL=	"JHEP",
	VOLUME=	"10",
	YEAR=	"2018",
	PAGES=	"094",
	DOI=	"10.1007/JHEP10(2018)094",
	NOTE=	"[Erratum: \DOI{10.1007/JHEP03(2019)063}]",
	EPRINT=	"1804.04474",
	ARCHIVEPREFIX=	"arXiv",
	PRIMARYCLASS=	"hep-ph",
}

@TECHREPORT{CMS-PAS-HIN-25-005,
	AUTHOR=	"{CMS Collaboration}",
	TITLE=	"{Event activity dependence of relative Y(nS) production in pPb collisions at 8.16 TeV}",
	URL=	"https://cds.cern.ch/record/2961526",
	TYPE=	"CMS Physics Analysis Summary",
	NUMBER=	"CMS-PAS-HIN-25-005",
	YEAR=	"2026",
}

@ARTICLE{Waagaard:2025dsi,
	AUTHOR=	"Waagaard, E. and Bruce, R. and Alemany Fernandez, R. and Bartosik, H. and Jowett, J. M. and Triantafyllou, N.",
	TITLE=	"{Charting the luminosity capabilities of the {CERN} {Large Hadron Collider} with various nuclear species}",
	EPRINT=	"2508.19653",
	ARCHIVEPREFIX=	"arXiv",
	PRIMARYCLASS=	"physics.acc-ph",
	DOI=	"10.1016/j.nima.2025.171118",
	JOURNAL=	"Nucl. Instrum. Meth. A",
	VOLUME=	"1083",
	PAGES=	"171118",
	YEAR=	"2026",
}

@INPROCEEDINGS{Brewer:2021kiv,
	AUTHOR=	"Brewer, Jasmine and Mazeliauskas, Aleksas and van der Schee, Wilke",
	TITLE=	"{Opportunities of \OO and p{O} collisions at the {LHC}}",
	BOOKTITLE=	"Opportunities of OO and pO collisions at the LHC",
	EPRINT=	"2103.01939",
	ARCHIVEPREFIX=	"arXiv",
	PRIMARYCLASS=	"hep-ph",
	REPORTNUMBER=	"CERN-TH-2021-028",
	YEAR=	"2021",
}

@UNPUBLISHED{ALICE:2025luc,
	AUTHOR=	"{ALICE Collaboration}",
	TITLE=	"{Evidence of nuclear geometry-driven anisotropic flow in \OO and \NeNe collisions at $\sqrtsNN = 5.36\TeV$}",
	EPRINT=	"2509.06428",
	ARCHIVEPREFIX=	"arXiv",
	PRIMARYCLASS=	"nucl-ex",
	REPORTNUMBER=	"CERN-EP-2025-203",
	YEAR=	"2025",
	NOTE=	"{Submitted to \textit{Phys. Rev. Lett.}}",
}

@ARTICLE{ATLAS:2025nnt,
	AUTHOR=	"{ATLAS Collaboration}",
	TITLE=	"{Measurement of the azimuthal anisotropy of charged particles in $\sqrtsNN=5.36\TeV$ $^{16}\mathrm{O}+{}^{16}\mathrm{O}$ and $^{20}\mathrm{Ne}+{}^{20}\mathrm{Ne}$ collisions with the {ATLAS} detector}",
	EPRINT=	"2509.05171",
	ARCHIVEPREFIX=	"arXiv",
	PRIMARYCLASS=	"nucl-ex",
	REPORTNUMBER=	"CERN-EP-2025-200",
	DOI=	"10.1103/xqxz-8bhf",
	JOURNAL=	"Phys. Rev. C",
	VOLUME=	"113",
	PAGES=	"045205",
	YEAR=	"2026",
}

@UNPUBLISHED{CMS:2025tga,
	AUTHOR=	"{CMS Collaboration}",
	TITLE=	"{Observation of long-range collective flow in \OO and \NeNe collisions and implications for nuclear structure studies}",
	EPRINT=	"2510.02580",
	ARCHIVEPREFIX=	"arXiv",
	PRIMARYCLASS=	"nucl-ex",
	REPORTNUMBER=	"CMS-HIN-25-009, CERN-EP-2025-222",
	YEAR=	"2025",
	NOTE=	"{Accepted by \textit{Phys. Rev. Lett.}}",
}

@ARTICLE{CMS:2025bta,
	AUTHOR=	"{CMS Collaboration}",
	TITLE=	"{Observation of suppressed charged-particle production in ultrarelativistic oxygen-oxygen collisions}",
	EPRINT=	"2510.09864",
	ARCHIVEPREFIX=	"arXiv",
	PRIMARYCLASS=	"nucl-ex",
	REPORTNUMBER=	"CMS-HIN-25-008, CERN-EP-2025-226",
	DOI=	"10.1103/89sf-9t1x",
	JOURNAL=	"Phys. Rev. Lett.",
	VOLUME=	"136",
	PAGES=	"162301",
	YEAR=	"2026",
}

@UNPUBLISHED{CMS:2026qef,
	AUTHOR=	"{CMS Collaboration}",
	TITLE=	"{System-size dependence of charged-particle suppression in ultrarelativistic nucleus-nucleus collisions}",
	EPRINT=	"2602.21325",
	ARCHIVEPREFIX=	"arXiv",
	PRIMARYCLASS=	"nucl-ex",
	REPORTNUMBER=	"CMS-HIN-25-014, CERN-EP-2026-032",
	YEAR=	"2026",
	NOTE=	"{Accepted by \textit{Phys. Lett. B}}",
}

@UNPUBLISHED{LHCb:2026gwz,
	AUTHOR=	"{LHCb Collaboration}",
	TITLE=	"{Study of nuclear effects on charm production in light ion collisions}",
	EPRINT=	"2605.27273",
	ARCHIVEPREFIX=	"arXiv",
	PRIMARYCLASS=	"nucl-ex",
	REPORTNUMBER=	"LHCb-PAPER-2026-008, CERN-EP-2026-141",
	YEAR=	"2026",
	NOTE=	"{Preprint}",
}

@UNPUBLISHED{CMS:2026sai,
	AUTHOR=	"{CMS Collaboration}",
	TITLE=	"{Centrality dependence of charged-hadron pseudorapidity distributions in oxygen-oxygen collisions at $\sqrtsNN=5.36\TeV$}",
	EPRINT=	"2606.02285",
	ARCHIVEPREFIX=	"arXiv",
	PRIMARYCLASS=	"nucl-ex",
	REPORTNUMBER=	"CMS-HIN-25-010, CERN-EP-2026-153",
	YEAR=	"2026",
	NOTE=	"{Submitted to \textit{Phys. Lett. B}}",
}

@UNPUBLISHED{ALICE:2026zck,
	AUTHOR=	"{ALICE Collaboration}",
	TITLE=	"{Evidence for parton energy loss in oxygen-oxygen collisions at $\sqrtsNN=5.36\TeV$}",
	EPRINT=	"2606.19967",
	ARCHIVEPREFIX=	"arXiv",
	PRIMARYCLASS=	"nucl-ex",
	REPORTNUMBER=	"CERN-EP-2026-171",
	YEAR=	"2026",
	NOTE=	"{Submitted to \textit{Phys. Rev. Lett.}}",
}

@UNPUBLISHED{ATLAS:2026zgq,
	AUTHOR=	"{ATLAS Collaboration}",
	TITLE=	"{Measurements of charged-particle pseudorapidity and transverse momentum distributions in \OO and \NeNe collisions at $\sqrtsNN=5.36\TeV$ with the {ATLAS} detector}",
	EPRINT=	"2606.20257",
	ARCHIVEPREFIX=	"arXiv",
	PRIMARYCLASS=	"nucl-ex",
	REPORTNUMBER=	"CERN-EP-2026-154",
	YEAR=	"2026",
	NOTE=	"{Submitted to \textit{Phys. Rev. C}}",
}

@UNPUBLISHED{ATLAS:2026fos,
	AUTHOR=	"{ATLAS Collaboration}",
	TITLE=	"{Observation of centrality-dependent dijet transverse momentum imbalance in \OO\ and \NeNe\ collisions at $\sqrtsNN=5.36\TeV$ with the {ATLAS} detector}",
	EPRINT=	"2606.20463",
	ARCHIVEPREFIX=	"arXiv",
	PRIMARYCLASS=	"nucl-ex",
	REPORTNUMBER=	"CERN-EP-2026-178",
	YEAR=	"2026",
	NOTE=	"{Submitted to \textit{Phys. Rev. Lett.}}",
}

@UNPUBLISHED{CMS:2025rzq,
	AUTHOR=	"{CMS Collaboration}",
	TITLE=	"{Luminosity measurement for lead-lead collisions at $\sqrtsNN = 5.02\TeV$ in 2015 and 2018 at {CMS}}",
	EPRINT=	"2503.03946",
	ARCHIVEPREFIX=	"arXiv",
	PRIMARYCLASS=	"hep-ex",
	REPORTNUMBER=	"CMS-LUM-20-002, CERN-EP-2025-018",
	YEAR=	"2025",
	NOTE=	"Submitted to \textit{Eur. Phys. J. C}",
}

@UNPUBLISHED{CMS:2026soy,
	AUTHOR=	"{CMS Collaboration}",
	TITLE=	"{Observation of nuclear suppression in coherent \PgUa\ photoproduction off heavy nuclei at the LHC}",
	EPRINT=	"2604.05814",
	ARCHIVEPREFIX=	"arXiv",
	PRIMARYCLASS=	"nucl-ex",
	REPORTNUMBER=	"CMS-HIN-24-013, CERN-EP-2026-050",
	YEAR=	"2026",
	NOTE=	"{Submitted to \textit{Phys. Rev. Lett.}}",
}

@ARTICLE{CMS:2017ycw,
	AUTHOR=	"{CMS Collaboration}",
	TITLE=	"{Suppression of excited \PgU states relative to the ground state in \PbPb collisions at $\sqrtsNN = 5.02\TeV$}",
	EPRINT=	"1706.05984",
	ARCHIVEPREFIX=	"arXiv",
	PRIMARYCLASS=	"hep-ex",
	REPORTNUMBER=	"CMS-HIN-16-008, CERN-EP-2017-080",
	DOI=	"10.1103/PhysRevLett.120.142301",
	JOURNAL=	"Phys. Rev. Lett.",
	VOLUME=	"120",
	PAGES=	"142301",
	YEAR=	"2018",
}

@MISC{hepdata,
	AUTHOR=	"{CMS Collaboration}",
	TITLE=	"{{HEPData} record for this analysis}",
	DOI=	"10.17182/hepdata.180655",
	YEAR=	"2026",
}

@ARTICLE{CMS:2008xjf,
	AUTHOR=	"{CMS Collaboration}",
	TITLE=	"{The {CMS} experiment at the {CERN} {LHC}}",
	DOI=	"10.1088/1748-0221/3/08/S08004",
	JOURNAL=	"JINST",
	VOLUME=	"3",
	PAGES=	"S08004",
	YEAR=	"2008",
}

@ARTICLE{CMS:2023gfb,
	AUTHOR=	"{CMS Collaboration}",
	TITLE=	"{Development of the {CMS} detector for the {CERN} {LHC} Run 3}",
	EPRINT=	"2309.05466",
	ARCHIVEPREFIX=	"arXiv",
	PRIMARYCLASS=	"physics.ins-det",
	REPORTNUMBER=	"CMS-PRF-21-001, CERN-EP-2023-136",
	DOI=	"10.1088/1748-0221/19/05/P05064",
	JOURNAL=	"JINST",
	VOLUME=	"19",
	PAGES=	"P05064",
	YEAR=	"2024",
}

@ARTICLE{CMS:2020cmk,
	AUTHOR=	"{CMS Collaboration}",
	TITLE=	"{Performance of the {CMS} level-1 trigger in \pp collisions at $\sqrt{s} = 13\TeV$}",
	JOURNAL=	"JINST",
	VOLUME=	"15",
	PAGES=	"P10017",
	YEAR=	"2020",
	DOI=	"10.1088/1748-0221/15/10/P10017",
	EPRINT=	"2006.10165",
	ARCHIVEPREFIX=	"arXiv",
	PRIMARYCLASS=	"hep-ex",
	REPORTNUMBER=	"CMS-TRG-17-001, CERN-EP-2020-065",
}

@ARTICLE{CMS:2016ngn,
	AUTHOR=	"{CMS Collaboration}",
	TITLE=	"{The {CMS} trigger system}",
	JOURNAL=	"JINST",
	VOLUME=	"12",
	PAGES=	"P01020",
	DOI=	"10.1088/1748-0221/12/01/P01020",
	YEAR=	"2017",
	EPRINT=	"1609.02366",
	ARCHIVEPREFIX=	"arXiv",
	PRIMARYCLASS=	"physics.ins-det",
	REPORTNUMBER=	"CMS-TRG-12-001, CERN-EP-2016-160",
}

@ARTICLE{CMS:2015xaf,
	AUTHOR=	"{CMS Collaboration}",
	TITLE=	"{Performance of electron reconstruction and selection with the {CMS} detector in \pp collisions at $\sqrt{s} = 8\TeV$}",
	EPRINT=	"1502.02701",
	ARCHIVEPREFIX=	"arXiv",
	PRIMARYCLASS=	"physics.ins-det",
	REPORTNUMBER=	"CMS-EGM-13-001, CERN-PH-EP-2015-004",
	DOI=	"10.1088/1748-0221/10/06/P06005",
	JOURNAL=	"JINST",
	VOLUME=	"10",
	PAGES=	"P06005",
	YEAR=	"2015",
}

@ARTICLE{CMS:2018rym,
	AUTHOR=	"{CMS Collaboration}",
	TITLE=	"{Performance of the {CMS} muon detector and muon reconstruction with \pp collisions at $\sqrt{s} = 13\TeV$}",
	EPRINT=	"1804.04528",
	ARCHIVEPREFIX=	"arXiv",
	PRIMARYCLASS=	"physics.ins-det",
	REPORTNUMBER=	"CMS-MUO-16-001, CERN-EP-2018-058",
	DOI=	"10.1088/1748-0221/13/06/P06015",
	JOURNAL=	"JINST",
	VOLUME=	"13",
	PAGES=	"P06015",
	YEAR=	"2018",
}

@ARTICLE{CMS:2015myp,
	AUTHOR=	"{CMS Collaboration}",
	TITLE=	"{Performance of photon reconstruction and identification with the {CMS} detector in \pp collisions at $\sqrt{s} = 8\TeV$}",
	EPRINT=	"1502.02702",
	ARCHIVEPREFIX=	"arXiv",
	PRIMARYCLASS=	"physics.ins-det",
	REPORTNUMBER=	"CMS-EGM-14-001, CERN-PH-EP-2015-006",
	DOI=	"10.1088/1748-0221/10/08/P08010",
	JOURNAL=	"JINST",
	VOLUME=	"10",
	PAGES=	"P08010",
	YEAR=	"2015",
}

@ARTICLE{CMS:2014pgm,
	AUTHOR=	"{CMS Collaboration}",
	TITLE=	"{Description and performance of track and primary-vertex reconstruction with the {CMS} tracker}",
	EPRINT=	"1405.6569",
	ARCHIVEPREFIX=	"arXiv",
	PRIMARYCLASS=	"physics.ins-det",
	REPORTNUMBER=	"CMS-TRK-11-001, CERN-PH-EP-2014-070",
	DOI=	"10.1088/1748-0221/9/10/P10009",
	JOURNAL=	"JINST",
	VOLUME=	"9",
	PAGES=	"P10009",
	YEAR=	"2014",
}

@TECHREPORT{CMS-TDR-15-02,
	AUTHOR=	"{CMS Collaboration}",
	TITLE=	"Technical proposal for the {Phase-II} upgrade of the {Compact Muon Solenoid}",
	URL=	"http://cds.cern.ch/record/2020886",
	TYPE=	"CMS Technical Proposal",
	NUMBER=	"CERN-LHCC-2015-010, CMS-TDR-15-02",
	INSTITUTION=	"CERN",
	YEAR=	"2015",
}

@ARTICLE{CMS:2017yfk,
	AUTHOR=	"{CMS Collaboration}",
	TITLE=	"{Particle-flow reconstruction and global event description with the {CMS} detector}",
	JOURNAL=	"JINST",
	VOLUME=	"12",
	YEAR=	"2017",
	PAGES=	"P10003",
	DOI=	"10.1088/1748-0221/12/10/P10003",
	EPRINT=	"1706.04965",
	ARCHIVEPREFIX=	"arXiv",
	PRIMARYCLASS=	"physics.ins-det",
	REPORTNUMBER=	"CMS-PRF-14-001, CERN-EP-2017-110",
}

@ARTICLE{CMS:2024qgs,
	AUTHOR=	"{CMS Collaboration}",
	TITLE=	"{Performance of {CMS} muon reconstruction from \pp to heavy ion collisions}",
	EPRINT=	"2404.17377",
	ARCHIVEPREFIX=	"arXiv",
	PRIMARYCLASS=	"hep-ex",
	REPORTNUMBER=	"CMS-MUO-21-001, CERN-EP-2024-040",
	DOI=	"10.1088/1748-0221/19/09/P09012",
	JOURNAL=	"JINST",
	VOLUME=	"19",
	PAGES=	"P09012",
	YEAR=	"2024",
}

@ARTICLE{Bierlich:2022pfr,
	AUTHOR=	"Bierlich, Christian and others",
	TITLE=	"{A comprehensive guide to the physics and usage of {\PYTHIA} 8.3}",
	EPRINT=	"2203.11601",
	ARCHIVEPREFIX=	"arXiv",
	PRIMARYCLASS=	"hep-ph",
	REPORTNUMBER=	"LU-TP 22-16, MCNET-22-04, FERMILAB-PUB-22-227-SCD",
	DOI=	"10.21468/SciPostPhysCodeb.8",
	JOURNAL=	"SciPost Phys. Codeb.",
	VOLUME=	"2022",
	PAGES=	"8",
	YEAR=	"2022",
}

@ARTICLE{CMS:2019csb,
	AUTHOR=	"{CMS Collaboration}",
	TITLE=	"{Extraction and validation of a new set of {CMS} {\PYTHIA8} tunes from underlying-event measurements}",
	EPRINT=	"1903.12179",
	ARCHIVEPREFIX=	"arXiv",
	PRIMARYCLASS=	"hep-ex",
	REPORTNUMBER=	"CMS-GEN-17-001, CERN-EP-2019-007",
	DOI=	"10.1140/epjc/s10052-019-7499-4",
	JOURNAL=	"Eur. Phys. J. C",
	VOLUME=	"80",
	PAGES=	"4",
	YEAR=	"2020",
}

@ARTICLE{CMS:UpsPolarization,
	AUTHOR=	"{CMS Collaboration}",
	TITLE=	"{Measurement of the \PgUa, \PgUb, and \PgUc polarizations in \pp collisions at $\sqrt{s} = 7\TeV$}",
	EPRINT=	"1209.2922",
	ARCHIVEPREFIX=	"arXiv",
	PRIMARYCLASS=	"hep-ex",
	REPORTNUMBER=	"CMS-BPH-11-023, CERN-EP-2012-205",
	DOI=	"10.1103/PhysRevLett.110.081802",
	JOURNAL=	"Phys. Rev. Lett.",
	VOLUME=	"110",
	PAGES=	"081802",
	YEAR=	"2013",
}

@ARTICLE{CMS:UpsPolarizationMult,
	AUTHOR=	"{CMS Collaboration}",
	TITLE=	"{$\PgU(n\text{S})$ polarizations versus particle multiplicity in \pp collisions at $\sqrt{s} = 7\TeV$}",
	EPRINT=	"1603.02913",
	ARCHIVEPREFIX=	"arXiv",
	PRIMARYCLASS=	"hep-ex",
	REPORTNUMBER=	"CMS-HIN-15-003, CERN-EP-2015-310",
	DOI=	"10.1016/j.physletb.2016.07.065",
	JOURNAL=	"Phys. Lett. B",
	VOLUME=	"761",
	PAGES=	"31",
	YEAR=	"2016",
}

@ARTICLE{LHCb:UpsPolarization,
	AUTHOR=	"{LHCb Collaboration}",
	TITLE=	"{Measurement of the \PgU polarizations in \pp collisions at $\sqrt{s} = 7$ and $8\TeV$}",
	EPRINT=	"1709.01301",
	ARCHIVEPREFIX=	"arXiv",
	PRIMARYCLASS=	"hep-ex",
	REPORTNUMBER=	"LHCb-PAPER-2017-028, CERN-EP-2017-210",
	DOI=	"10.1007/JHEP12(2017)110",
	JOURNAL=	"JHEP",
	VOLUME=	"12",
	PAGES=	"110",
	YEAR=	"2017",
}

@ARTICLE{Acharya:2021PLB,
	AUTHOR=	"{ALICE Collaboration}",
	TITLE=	"{First measurement of quarkonium polarization in nuclear collisions at the {LHC}}",
	JOURNAL=	"Phys. Lett. B",
	VOLUME=	"815",
	PAGES=	"136146",
	YEAR=	"2021",
	DOI=	"10.1016/j.physletb.2021.136146",
	EPRINT=	"2005.11128",
	ARCHIVEPREFIX=	"arXiv",
	PRIMARYCLASS=	"nucl-ex",
	REPORTNUMBER=	"CERN-EP-2020-092",
}

@ARTICLE{CMS:2020efs,
	AUTHOR=	"{CMS Collaboration}",
	TITLE=	"{Measurement of the azimuthal anisotropy of \PgUa and \PgUb mesons in \PbPb collisions at $\sqrtsNN = 5.02\TeV$}",
	EPRINT=	"2006.07707",
	ARCHIVEPREFIX=	"arXiv",
	PRIMARYCLASS=	"hep-ex",
	REPORTNUMBER=	"CMS-HIN-19-002, CERN-EP-2020-039",
	DOI=	"10.1016/j.physletb.2021.136385",
	JOURNAL=	"Phys. Lett. B",
	VOLUME=	"819",
	PAGES=	"136385",
	YEAR=	"2021",
}

@ARTICLE{Wang:1991hta,
	AUTHOR=	"Wang, Xin-Nian and Gyulassy, Miklos",
	TITLE=	"{{HIJING}: a {Monte Carlo} model for multiple jet production in \pp, \pPb and \AAa collisions}",
	REPORTNUMBER=	"LBL-31036",
	DOI=	"10.1103/PhysRevD.44.3501",
	JOURNAL=	"Phys. Rev. D",
	VOLUME=	"44",
	PAGES=	"3501",
	YEAR=	"1991",
}

@ARTICLE{Agostinelli:2002hh,
	AUTHOR=	"Agostinelli, S. and others",
	COLLABORATION=	"GEANT4",
	TITLE=	"{{\GEANTfour}---a simulation toolkit}",
	JOURNAL=	"Nucl. Instrum. Meth. A",
	VOLUME=	"506",
	YEAR=	"2003",
	PAGES=	"250",
	DOI=	"10.1016/S0168-9002(03)01368-8",
}

@PHDTHESIS{Oreglia:1980cs,
	AUTHOR=	"Oreglia, Mark Joseph",
	TITLE=	"A study of the reactions $\psi^\prime \to \gamma \gamma \psi$",
	SCHOOL=	"Stanford University",
	YEAR=	"1980",
	NOTE=	"{SLAC} Report {SLAC-R-236}",
	URL=	"https://www.slac.stanford.edu/pubs/slacreports/reports13/slac-r-236.pdf",
}

@PHDTHESIS{Gaiser:1982yw,
	AUTHOR=	"Gaiser, John Erthal",
	TITLE=	"{Charmonium spectroscopy from radiative decays of the $\mathrm{J}/\psi$ and $\psi^\prime$}",
	NOTE=	"{SLAC} Report {SLAC-R-255}",
	SCHOOL=	"Stanford University",
	YEAR=	"1982",
	URL=	"http://www-public.slac.stanford.edu/sciDoc/docMeta.aspx?slacPubNumber=slac-r-255.html",
}

@ARTICLE{Loizides:2025ule,
	AUTHOR=	"Loizides, Constantin",
	TITLE=	"{Glauber predictions for oxygen and neon collisions at energies available at the CERN Large Hadron Collider}",
	EPRINT=	"2507.05853",
	ARCHIVEPREFIX=	"arXiv",
	PRIMARYCLASS=	"nucl-th",
	DOI=	"10.1103/mkp8-zgxh",
	JOURNAL=	"Phys. Rev. C",
	VOLUME=	"113",
	PAGES=	"014914",
	YEAR=	"2026",
}

@ARTICLE{CMS:2020fae,
	AUTHOR=	"{CMS Collaboration}",
	TITLE=	"{Investigation into the event-activity dependence of \PGUPnS relative production in \pp collisions at $\sqrt{s} = 7\TeV$}",
	EPRINT=	"2007.04277",
	ARCHIVEPREFIX=	"arXiv",
	PRIMARYCLASS=	"hep-ex",
	REPORTNUMBER=	"CMS-BPH-14-009, CERN-EP-2020-075",
	DOI=	"10.1007/JHEP11(2020)001",
	JOURNAL=	"JHEP",
	VOLUME=	"11",
	PAGES=	"001",
	YEAR=	"2020",
}

@ARTICLE{Kim:2022lgu,
	AUTHOR=	"Kim, Junlee and Seo, Jinjoo and Hong, Byungsik and Hong, Juhee and Kim, Eun-Joo and Kim, Yongsun and Kweon, MinJung and Lee, Su Houng and Lim, Sanghoon and Park, Jaebeom",
	TITLE=	"{Model study on \PGUPnS modification in small collision systems}",
	EPRINT=	"2209.12303",
	ARCHIVEPREFIX=	"arXiv",
	PRIMARYCLASS=	"nucl-th",
	DOI=	"10.1103/PhysRevC.107.054905",
	JOURNAL=	"Phys. Rev. C",
	VOLUME=	"107",
	PAGES=	"054905",
	YEAR=	"2023",
}
